\documentclass[aps,pra,showpacs,preprint,onecolumn]{revtex4-1}
\usepackage{hyperref}
\usepackage{xcolor}
\hypersetup{
	colorlinks,
	linkcolor={red!50!black},
	citecolor={blue!50!black},
	urlcolor={blue!80!black}}
\usepackage{amsmath}
\usepackage{amsfonts}
\usepackage{amssymb}
\usepackage{graphicx}
\usepackage{stackrel}

\newcommand{\ket}[1]{|#1\rangle}                      		% Ket Dirac's notation %

\begin{document}
\title{Continuous-Variable Entangled States of Light carrying Orbital Angular Momentum}

\author{A. Pecoraro$^{1,2}$, F. Cardano$^{2}$, L. Marrucci$^{2}$ and A. Porzio$^{1}$\footnote{Corresponding author: alberto.porzio@spin.cnr.it}}
\affiliation{$^1$ SPIN -- CNR, Napoli, Complesso Universitario di Monte Sant'Angelo, via Cintia, 80126 Napoli, Italy}
\affiliation{$^2$ Dipartimento di Fisica, Universit\`a di Napoli Federico II, via Cintia, 80126 Napoli, Italy}
\date{\today}

\begin{abstract}
The orbital angular momentum of light, unlike spin, is an infinite-dimensional discrete variable and may hence offer enhanced performances for encoding, transmitting, and processing quantum information. Hitherto, this degree of freedom of light has been studied mainly in the context of quantum states with definite number of photons. On the other hand, field-quadrature continuous-variable quantum states of light allow implementing many important quantum protocols not accessible with photon-number states. Here, we realize a novel scheme based on a q-plate device for endowing a bipartite continuous-variable Gaussian entangled state with non-zero orbital angular momentum. We then apply a reconfigurable homodyne detector working directly with such non-zero orbital angular momentum modes in order to retrieve experimentally their entire quantum-state covariance matrix, thus providing a full characterisation of their quantum fluctuation properties. Our work is a step towards generating multipartite continuous-variable entanglement in a single optical beam.
\end{abstract}

\maketitle

\section{Introduction}

Accessing high-dimensional effective Hilbert spaces is crucial for implementing complex quantum information (QI) tasks. The continuous-variable (CV) encoding provides a convenient possible approach to this purpose, as it spans an inherently infinite Hilbert space and it naturally allows quantum communication and processing with larger alphabets \cite{Grosshans2002,Grosshans2003}. This approach is often investigated within quantum optical platforms, by using optical field quadratures as the variables.

Gaussian bipartite entanglement in optical CV QI is commonly generated by means of below-threshold optical parametric oscillators (OPOs), which exploit either spatial separation or polarization, i.e.\ spin angular momentum (SAM), as distinguishing degree-of-freedom (d.o.f.) \cite{Ou1992,Bowen2003,Laurat2005,PRL2009}. It would be, however, highly desirable to move towards CV \textit{multipartite} entanglement. This goal could be possibly sought by introducing additional discrete degrees of freedom with which labelling CVs,
that is by adopting what can be termed a ``hybrid discrete-continuous variable'' strategy to carrying out QI. Eventually, a single degree of freedom can be exploited, in case it is associated with multiple orthogonal modes. Multipartite CV entanglement has been for example pursued by exploiting time/frequency domains in pulsed systems \cite{Chalopin2011,Yokoyama2013,Roslund2014,Chen2014} or by combining the outcome of a number of single mode squeezing sources \cite{Yukawa2009}. These approaches may also lead to other interesting outcomes, such as CV hyper-entanglement \cite{dosSantos2009}.

Another possible route for implementing such a strategy is by using transverse modes of a single optical beam as additional discrete d.o.f. Transverse optical modes, and in particular those carrying orbital angular momentum (OAM), have been the subject of much research in quantum optics, recently. In particular, OAM can conveniently provide a large set of orthogonal modes within a single optical beam, as demonstrated in a number of experiments in which high-dimensional Hilbert spaces and entanglement have been achieved \cite{Erhard2018}. However, this research focused mainly on ``digital'' photon-number quantum optics \cite{Halina2017,Erhard2017}, while not much has been done hitherto by combining CV quantum optics and OAM.  

An indirect proof of Gaussian CV entanglement between OAM modes has been first reported by Lassen \textit{et al.} \cite{Lassen2009}, specifically by showing quantum squeezing of a first-order Hermite-Gaussian (HG) mode that corresponds to a balanced superposition of two opposite-OAM ($\pm 1$) Laguerre-Gaussian (LG) modes. A similar approach, i.e.\ the measurement of quadrature-squeezing of first-order HG modes, has been adopted by Liu \textit{et al.} \cite{Liu2014} to demonstrate experimentally the first hyper-entangled CV state. However, both these works demonstrated directly only the squeezing of the OAM-superposition HG modes. A complete characterization of OAM-mode CV entanglement, that for Gaussian states amounts to measuring the full covariance matrix of the two entangled modes \cite{Olivares2012}, has been hitherto missing.

In the work reported here, we endow a pair of cross-polarized CV entangled modes with non-zero OAM, thus realising multi-d.o.f. CV entanglement with the two entangled modes labelled by both SAM and OAM. Then, we use a novel spatial-mode-reconfigurable homodyne detector \cite{OAM_homodyne} to measure directly the quadratures of various linear combinations of the two entangled CVs, thus enabling us to reconstruct their full covariance matrix (CM). In this way, we provide what is the first complete and direct characterisation of such a SAM-OAM CV entanglement, to our knowledge. The paper also contains a detailed discussion on the operative procedure for data analysis that allows one checking the compatibility of the measured state with the state one expects from the generation method and source parameter as well as the influence of the SAM to OAM conversion.

The paper is organized as follows. In Sect.\ \ref{Sect:OAM} we descrive, briefly, how the OAM carrying entangled state is obtained starting from the entanglement source, a type--II OPO. Moreover, we discuss the main features one expects to detect in the resulting light field. In Sect.\ \ref{Sect:Data_State} we illustrate the data analysis procedure and the main experimental results. The following Sect.\ \ref{Sect:Physical_state} is devoted to a discussion of the compatibility of the measured state with the known generation and detection parameters. Finally, in Sect.\ \ref{Sect:Conclusions} some conclusions are drawn.

\section{Entanglement generation and OAM manipulation}
\label{Sect:OAM}

Optical beams for which photons carry a definite amount of OAM correspond to the so-called ``helical modes'' of light, i.e.\ paraxial light beams characterized by a helical phase factor 
$e^{im\varphi}$ where $\varphi$ is the azimuthal angle around the propagation axis $z$ and $m$ is an integer. An optical vortex with topological charge $m$ is then present on axis, where the phase is undefined and the light intensity vanishes (doughnut beams) \cite{Yao2011,Padgett2017}. In these modes, the ratio between OAM and energy fluxes along the optical axis is $m/\omega$, where $\omega$ is the optical frequency. In other words, each photon within the beam carries $m\hbar$ of OAM along the propagation direction. Quantum states $\ket m$, describing individual photons whose spatial structure is that of an helical mode, are orthogonal and span an infinite-dimensional Hilbert space.

\begin{figure}[!tph]
	\centering\includegraphics[width=0.38\textwidth]{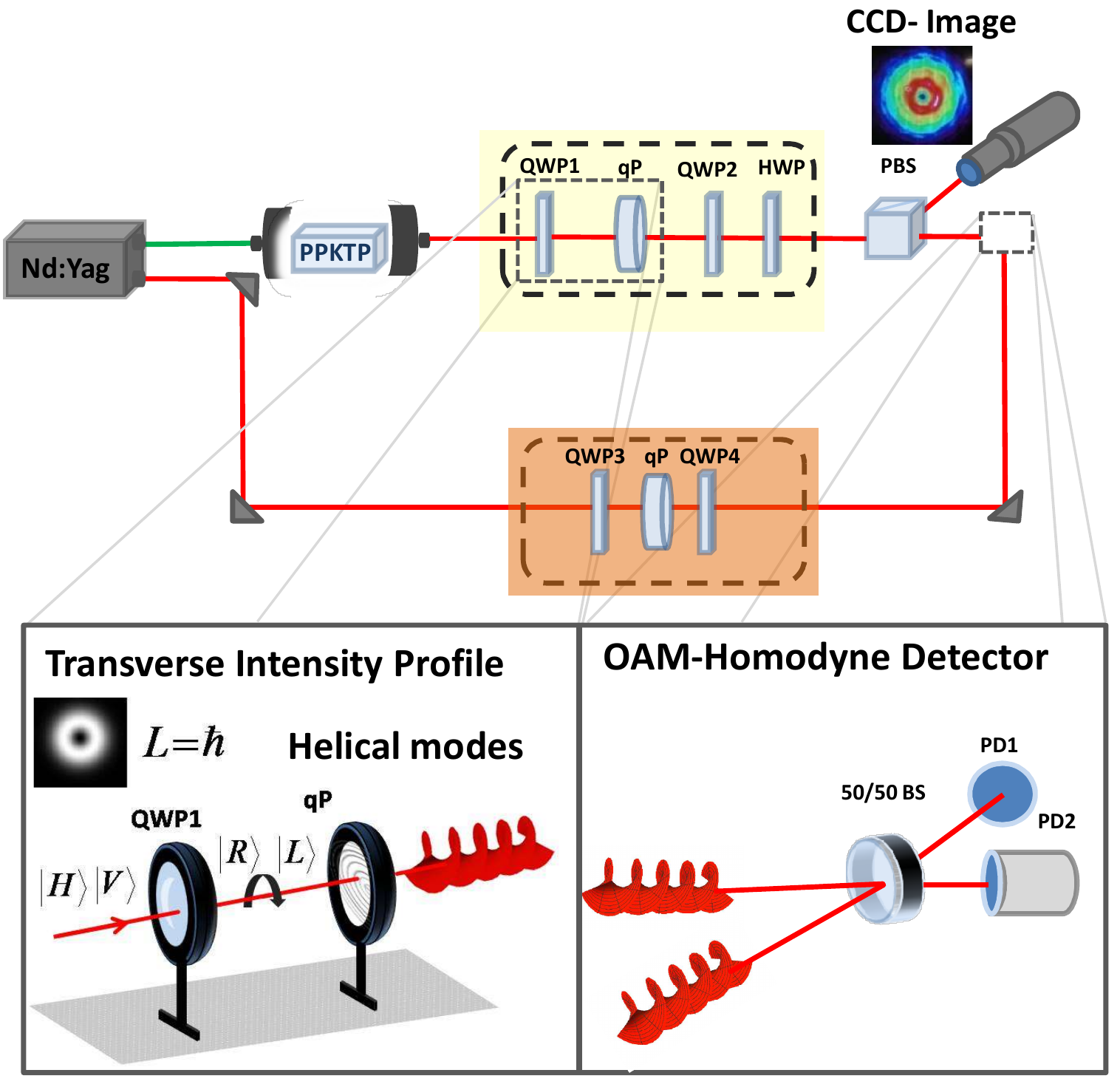}
	\caption{Schematic of the experimental set-up. An internally frequency-doubled Nd:Yag pumps an OPO based on a $\alpha$-cut PPKTP type-II crystal that generates a pair of frequency-degenerate orthogonally-polarized modes. The OPO \cite{APB2008} oscillation threshold is $\approx 70$ mW and it is operated at $\approx 70\%$
	of the threshold value. The yellow shadowed area
	(bottom-left inset)
	including two quarter-wave plates (QWP1 and QWP2), a $q$-plate (qP) and a half-wave plate (HWP), represents the optical set-up for manipulating the SAM and OAM d.o.f. of the entangled beam. A similar set of optical components
	is placed along the LO path for effective OAM homodyning (dark orange shadow). The bottom-right inset shows an enlarged view of the homodyne detector.}
	\label{Fig:Exp}
\end{figure}

In our experiment (see Fig.\ \ref{Fig:Exp} for the setup layout and some technical details), the initial quantum source is a standard type-II phase-matching OPO generating Gaussian bipartite entangled states \cite{IJQI2007}. The phase matching is adjusted to a degenerate condition \cite{APB2008}, so that the two beams have the same frequency, corresponding to a wavelength of 1064 nm. The entangled modes emerge from the OPO as two cross-linearly-polarized continuous-wave (CW) co-propagating beams, both having a Gaussian transverse profile (TEM$_{00}$), i.e.\ with vanishing OAM. Let us label as $H$ ($V$) the horizontal (vertical) linear polarization of the modes generated by the OPO.

OAM is imparted to the two modes generated by the OPO by using a device commonly known as ``$q$-plate'' (qP) \cite{Marrucci2006}. Essentially, a $q$-plate is a thin layer of birefringent liquid crystal, sandwiched between containing glasses, and whose optic axis is structured into a singular pattern with topological charge $q$, the latter being an integer or half-integer number. The total birefringent phase retardation $\delta$ can be controlled electrically \cite{Piccirillo2010}. Using a quantum notation, the optical action of a $q$-plate can be simply described by a unitary operator $\hat Q$ defined as follows:
\begin{align}
\hat Q\; k_{L,m} \hat{Q}^{\dagger} = \cos \frac{\delta}{2}\; k_{L,m} +i\sin \frac{\delta}{2}\; k_{R,m+2q} \nonumber \\
\hat Q\; k_{R,m} \hat{Q}^{\dagger} = \cos \frac{\delta}{2}\; k_{R,m} +i\sin \frac{\delta}{2}\; k_{L,m-2q}
\label{Eq:q-plate}
\end{align}
where $k$ is the generic electromagnetic mode operator and $ \left(A,m\right) $ labels a mode with OAM $m$ (in units of $\hbar$) and polarization $A=L, R$, with $L$ ($R$) standing for left (right) circular polarization. The radial structure of the beam is also affected by the qP, but we omit it in our notation for brevity (all modes having the same $|m|$ also have the same radial structure). From Eq.\ (\ref{Eq:q-plate}) it is seen that a qP couples circular polarization to OAM: for the optimal retardation $\delta=\pi$, a single circularly polarized beam passing through the qP reverses its polarization and acquires $\pm 2q$ quanta of OAM. In our experiment we wish to use the same qP for both beams generated by the OPO, which have orthogonal linear polarizations. Therefore, we first transform these two linear polarizations into circular, by using a quarter-wave plate (QWP1, in Fig.\ \ref{Fig:Exp}) whose axis are rotated by $45^{\circ}$ with respect to the $H/V$ basis
so that $k_{V} \xrightarrow{\mathtt{QWP1}} ik_{R}$ and $k_{H}\xrightarrow{\mathtt{QWP1}}k_{L}$ (with no OAM variation). Next, the cross-polarized pair of CV entangled beams, now having opposite circular polarizations, pass through the qP, thus inverting their polarization handedness and simultaneously acquiring opposite OAM values (see the bottom--left inset in Fig.\ \ref{Fig:Exp}). In our experiment the qP has $q=1/2$, so that the acquired OAM is $m=\pm 1$. At the output of the qP, the two entangled modes are still co-propagating and can now be distinguished both by polarization ($L/R$) and OAM ($\pm1$).

Let us now discuss the quantum fluctuation/correlation properties of these two modes.
A sub-threshold OPO is known to generate a pair of modes in a Gaussian quantum state (GS) \cite{Olivares2012}.
The quantum properties of such two-mode GS are described by the canonical field quadrature operators $R=\left(X_{a},Y_{a},X_{b},Y_{b}\right)$, where the subscripts $a$ and $b$ label the 
two modes and $X_{k}=\left(\hat{a}_{k}+\hat{a}_{k}^{\dagger}\right)/\sqrt{2}$ and $Y_{k}=\left(\hat{a}_{k}-\hat{a}_{k}^{\dagger}\right)/(i\sqrt{2})$, with $\hat{a}_{k}$ 
($\hat{a}_{k}^{\dagger}$) the annihilation (creation) operator. The GS itself is completely 
characterized by the vector of mean values for $ X_{k} $ and $ Y_{k} $ and, most importantly, by 
their CM. The CM $\sigma$ of a bipartite GS is a real symmetric and positive-definite $4\times4$ 
matrix whose elements are defined as $\sigma_{ij}=\frac{1}{2}\left\langle \left\{ 
R_{i,}R_{j}\right\} \right\rangle -\left\langle R_{i}\right\rangle \left\langle 
R_{j}\right\rangle$, with $\left\{ h,g\right\} =hg+gh$.

At the output of a type--II OPO crystal, given the symmetry of its Hamiltonian \cite{Zhang1999}, the state we expect is
a pure two--mode squeezed state characterized by a single squeezing parameter $r$. In principle, if a photon is born in one mode, its twin populates the other mode of the pair. This implies that the total energy of the state is symmetrically distributed between the two entangled modes.
The CM relative to such twin-beam state, in the basis $(X_a,Y_a,X_b,Y_b)$, is in the so-called standard form
\begin{equation}
\sigma_{S}=\left(\begin{array}{cccc}
m & 0 & c_{1} & 0\\
0 & m & 0 & c_{2}\\
c_{1} & 0 & n & 0\\
0 & c_{2} & 0 & n
\end{array}\right)
\label{Eq:twin-beam-pure-matrix}
\end{equation}
with $c_{1}=-c_{2}=\sinh(2r)$ and
$n=m=\cosh(2r)$, due to the symmetry of the Hamiltonian, and $\Delta \sigma = 1$ (pure state condition).
All the relevant criteria for classifying the bipartite-state quantum properties can be written in terms of the CM elements \cite{PRA2012,PS2013}, so that its knowledge allows deducing information about the separability or entanglement of the state.

The knowledge of the physical process that gives rise to the measured quantum state allows one to make certain ``a priori'' assumptions, to be verified once data are analyzed. The two OPO modes travel together, in a single beam, and all subsequent optical elements are not expected to affect their symmetry significantly.  Therefore,
any residual state asymmetry in the measurement outcome can be ascribed to experimental imperfections and statistical fluctuations of the working condition during the acquisition time. To be compatible with our generation and detection scheme, the measured matrix should hence descend from a pure symmetric twin-beam state which has undergone symmetric losses. As a consequence, we expect, at the end of the data analysis, a matrix compatible with the form of Eq.\ (\ref{Eq:twin-beam-pure-matrix}), representing a bipartite entangled state whose losses are compatible with the level of optical losses estimated \textit{off-line} for our set-up.

Let us now label the pair of modes at the OPO output as $(a,b)$ \cite{PRL2009}. The initial two-mode quantum state is $HV$ polarized and has vanishing OAM, so it can be denoted as $a_{H,0},b_{V,0}$. The GS fluctuation properties (as given by the CM) are known to be preserved under linear optical transformations. Therefore, the action of the QWP1 and qP yields the following transformation:
\begin{equation}
\left(a_{H,0},b_{V,0}\right) \xrightarrow{\mathtt{QWP1}+\mathtt{qP}}  \left(ia_{R,1},-b_{L,-1}\right).
\label{Eq:ab-qplate-transf}
\end{equation}
At this stage, the resulting bipartite state has acquired an additional d.o.f. for distinguishing the two sub-systems $a$ and $b$, namely OAM, in addition to polarization (or spin). Here polarization and OAM play the role of mode labels and so they are referred to as distinguishability d.o.f. We are interested in studying the mutual correlation properties in phase space, i.e.\ studying the fluctuation in the field quadratures, of these two modes.

\section{Data analysis and state characterization}
\label{Sect:Data_State}

In order to characterize the quantum properties of this multi-distinguishable CV entangled pair, we wish to apply the single homodyne scheme discussed in Ref.\ \cite{JOB2005}. This, in turn, requires to homodyne modes $a$, $b$ as well as the following auxiliary linear-combination modes:
\begin{equation}
c=\frac{a+b}{\sqrt{2}} \qquad d=\frac{a-b}{\sqrt{2}} \qquad
e=\frac{ia+b}{\sqrt{2}} \qquad f=\frac{ia-b}{\sqrt{2}}
\label{Eq:auxiliary}
\end{equation}
In Ref.\ \cite{PRL2009}, these auxiliary modes could be easily obtained from the $(a,b)$ pair by combination of wave-plates and polarizing beam-splitters (also exploiting the frequency-degeneracy of the two modes). However, we now have the additional OAM d.o.f. and the two modes $a$ and $b$ have opposite values of OAM, which makes the task of creating their linear combinations much harder. To overcome this problem, we exploit the qP polarization-control properties for generating the modes $c, d, e$, and $f$ at the output of the qP (see Fig.\ \ref{Fig:Exp}). First, by turning the QWP1 so that its fast/slow axes coincide with $H/V$, we have $a_{H} \xrightarrow{\mathtt{QWP1}} a_{H}$ and $b_{V} \xrightarrow{\mathtt{QWP1}} ib_{V}$. At the qP output we then obtain 
\begin{align}
a_{H,0}=\frac{a_{L,0} +a_{R,0}}{\sqrt{2}}\xrightarrow{\mathtt{qP}}  
i\frac{a_{L,-1}+ a_{R,1}}{\sqrt{2}} \nonumber \\
ib_{V,0}=\frac{b_{L,0}-b_{R,0} }{\sqrt{2}}\xrightarrow{\mathtt{qP}}  
i \frac{b_{R,1}- b_{L,-1}}{\sqrt{2}}
\label{Eq:qP-cd} 
\end{align}
Then, grouping for the same polarization and OAM state and neglecting global phase factors, we obtain $c_{R,1}=\frac{a_{R,1}+b_{R,1}}{\sqrt{2}}$ and $d_{L,-1}=\frac{a_{L,-1}-b_{L,-1}}{\sqrt{2}}$. $c$ and $d$ modes are assigned to two opposite OAM modes and circular polarizations. Similarly, by removing the QWP1 altogether, so that the $H/V$ linear-polarized entangled modes enter directly the qP, we obtain
\begin{align}
a_{H,0}=\frac{a_{L,0} +a_{R,0}}{\sqrt{2}}\xrightarrow{\mathtt{qP}}  
i\frac{a_{L,-1}+ a_{R,1}}{\sqrt{2}} \nonumber \\
b_{V,0}=-i\frac{b_{L,0}-b_{R,0} }{\sqrt{2}}\xrightarrow{\mathtt{qP}}  
 \frac{b_{R,1}- b_{L,-1}}{\sqrt{2}}.
\label{Eq:qP-ef} 
\end{align}
Hence, at the qP output we now have $e_{R,1}=\frac{ia_{R,1}+b_{R,1}}{\sqrt{2}}$ and $f_{L,1}=\frac{ia_{L,1}-b_{L,1}}{\sqrt{2}}$. 
i.e., modes $e$ and $f$, again in the circular polarization basis and with opposite OAM.

The characterization scheme requires that we homodyne these six modes. A homodyne detector relies on the interference of the optical signal under scrutiny with a strong (coherent)
local oscillator (LO). Thus, we have designed a LO branch that generates a coherent optical field in the $(H,\pm 1)$ mode, i.e.\ horizontally polarized and with OAM $=\pm 1$. The homodyne interference is therefore designed, for the first time to our knowledge, to take place directly in the OAM state. The $H$ polarization of the LO is selected because it is the working polarization of the homodyne beamsplitter (BS) \cite{OAM_homodyne}. Before the homodyne interference, a second QWP (QWP2 in Fig.\ \ref{Fig:Exp}) is hence used to revert the polarization state of any mode pair to be characterized from the $L/R$ to the $H/V$ basis, while leaving the OAM unchanged. QWP2 is oriented so that $\left( R,1,L,-1 \right)\xrightarrow{\mathtt{QWP2}}\left( V,1,H,-1 \right)$. A final half-wave plate (HWP) and a polarizing beam splitter (PBS) in front of the homodyne BS allows one to select which of the modes effectively reaches the
detection stage (see Fig.\ \ref{Fig:Exp}). A similar set-up, composed of two QWPs (QWP3 and QWP4 in Fig.\ \ref{Fig:Exp}) and a second qP tailors
the LO to the desired $\left[H, \pm1 \right]$ state. The homodyne visibility routinely obtained is $0.97 \pm 0.01$.

The overall detection efficiency includes photodiodes' quantum efficiency, homodyne visibility, and losses at the photodiodes' uncoated windows. The total collection efficiency 
has to take into account also the cavity output coupling and the qP transmission. All these factors lead to an overall collection efficiency of $0.52 \pm 0.03$. We note that the reduction in the collection efficiency due to the OAM endowing scheme (waveplates and qP) can be calculated by comparing the above value of $0.52 $ with the value of
$0.63 $ reported in Ref.\ \cite{PRA2012}. The additional losses introduced by the elements involved in the entanglement manipulation scheme (waveplates and qP) are, then, $0.175$. These losses could probably be reduced to $< 0.05$ by simple technical improvements, such as adding anti-reflection coatings to the qP. Despite the losses, the measured CM corresponds to an entanglement level that is sufficient to realize a CV quantum teleportation protocol \cite{Pirandola2006}.

Experimentally we have acquired quadrature traces for the six modes $(a,b,c,d,e,f)$. Data have been analyzed to retrieve the state CM. 

There is a physical constraint on the validity of this single homodyne method. The system must be stationary. The source has to produce the same state, within the experimental errors, for the whole acquisition time windows. This is not just a technical request. The idea of measuring auxiliary modes requires that the expectation values and variances for the quadrature operators of the original pair of modes are \textit{constant} during the acquisition time needed for scanning all the involved modes.

To have a continuous monitoring of the stationary requirement, the data acquisition system panel calculates, for each homodyne trace, the average photon number of each of the six measured modes. This quantity is proportional to the variance of data acquired over a 2$\pi$ scan of the generalized quadrature, independently of the type of Gaussian state being measured. This simple relation allows one to roughly monitor the stability of the state during the whole acquisition process. The six modes required for reconstructing the CM are saved only if they show the same average photon numbers, within a few percent tolerance.

\subsection{Gaussianity test}
The CM is in a one--to--one correspondence with the density matrix of the state if and only if the state is Gaussian, where Gaussianity is defined by the Gaussian condition on its Wigner function \cite{Olivares2012}. A Gaussian Wigner function implies that homodyne data recorded at a fixed angle, physically corresponding to the marginal distribution related to a single generalized quadrature operator, are Gaussian distributed \cite{Olivares2012,PRA2009}. At the same time, the OPO Hamiltonian, bi-linear in the field-mode operators, preserves the Gaussianity of the underlying stochastic processes \cite{Simon1987,Marian1993} even if it has been proven that, close to the oscillation threshold, some non-Gaussian effects, connected to usually negligible nonlinear terms that come into play, may appear \cite{OPEX2005,PRA2010}. The data presented in this paper actually correspond to an OPO working at $0.7\times \mathtt{P_{th}}$, a regime that is surely Gaussian.

\begin{figure}[!tph]
	\centering\includegraphics[width=0.48\textwidth]{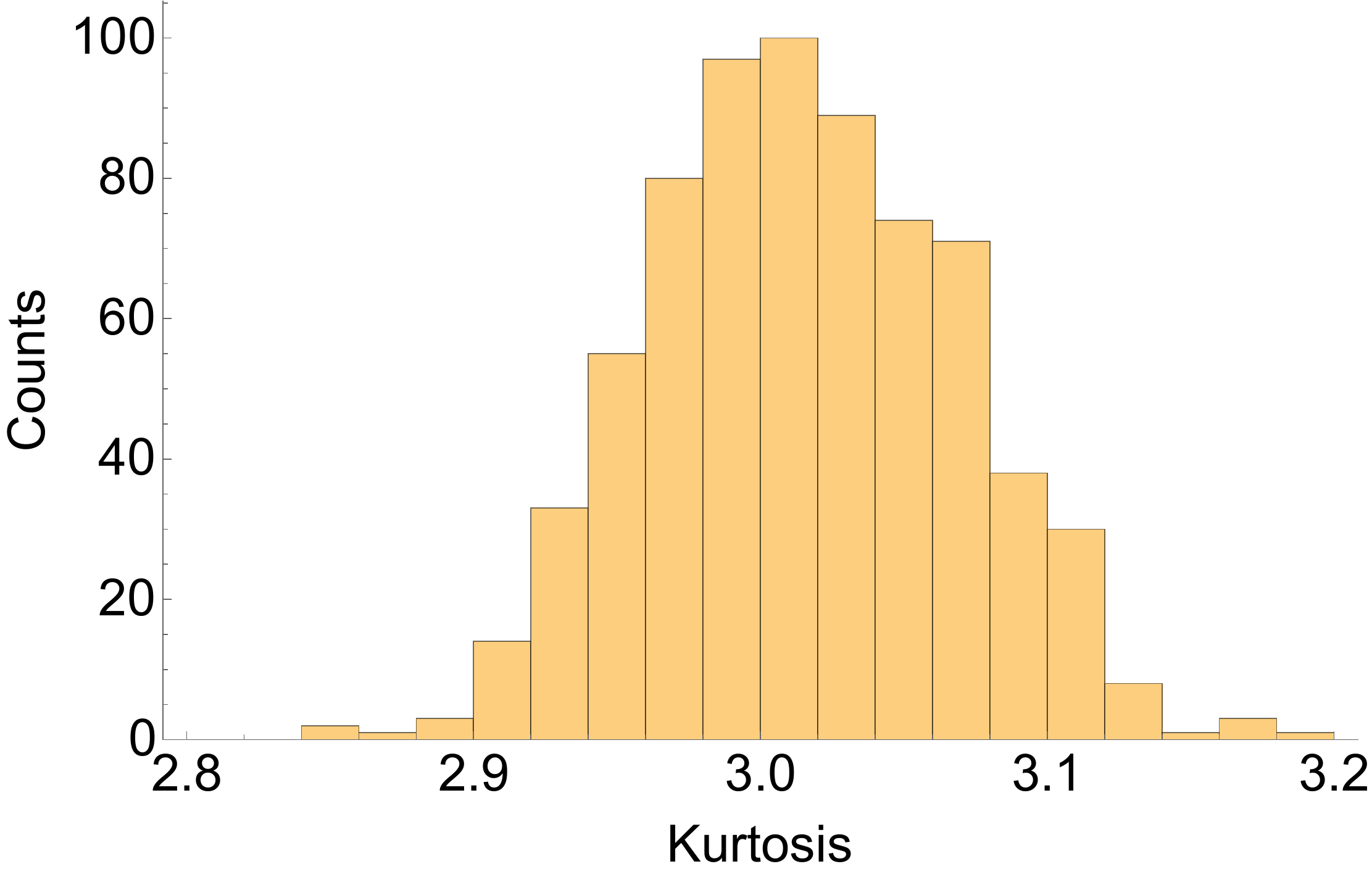}
	\caption{Histogram of the values obtained for the kurtosis relative to all the 700 quadrature marginal distributions used in the data analysis. Each data set, corresponding to a single detected mode, is partitioned into 100 phase bins each containing a marginal distribution of $10000$ points. The mean of the kurtosis distribution is $3.01$ with a standard deviation of $0.05$, thus confirming that the quadrature marginal distributions are Gaussian and so are the corresponding field states.}
	\label{Fig:Kurtosis_Histo}
\end{figure}

To experimentally verify this assumption, each single-mode data set undergoes a Gaussianity test. The kurtosis, i.e.\ the fourth-order moment, is calculated for all the marginal distributions obtained dividing each $2 \pi$ scan into 100 phase intervals. In Fig.\ \ref{Fig:Kurtosis_Histo} we report the histogram of kurtosis calculated for all 700 marginal distributions corresponding to the whole data set recorded for the six modes used for retrieving the CM reported below, plus the shot-noise acquisition. As it can be seen, the average value is $3.01$ with a standard deviation of $0.05$ where $3$ is the kurtosis expected for a Gaussian distributed variable. We note that in Ref.\ \cite{OPEX2005} systematic deviations from Gaussianity with kurtosis as high as $\sim 3.5$ have been found.

\subsection{Experimental Covariance Matrix}
\begin{figure}[tph]
	\centering\includegraphics[width=0.3\textwidth]{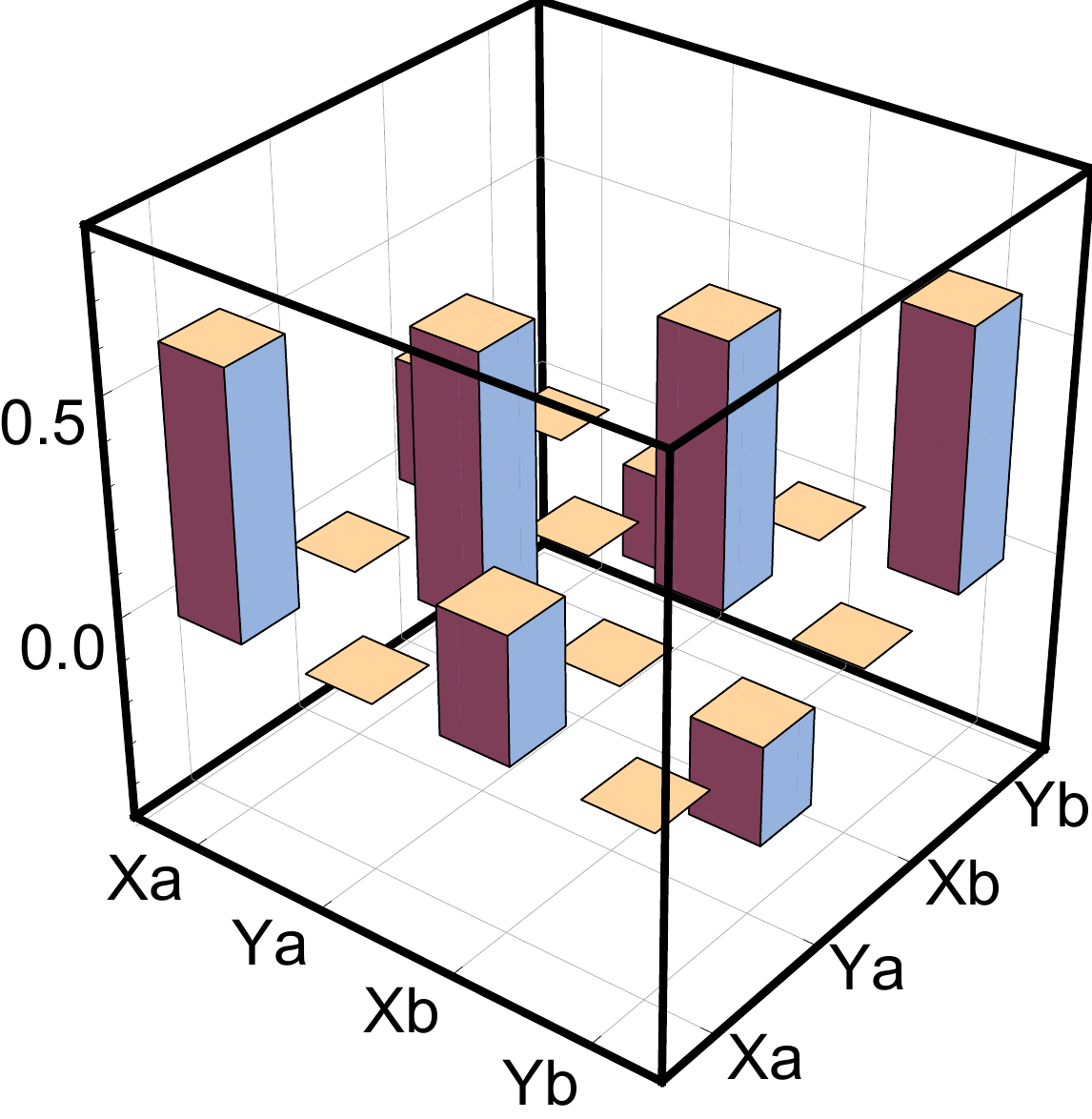}
	\caption{Graphic representation of the covariance matrix given in Eq.\ (\ref{Eq:CM-exp}).
	The non-zero elements outside the main diagonal are a signature of quantum correlation between
pairs of quadratures of modes carrying different OAM.}
	\label{Fig:Matrix}
\end{figure}

A typical obtained experimental covariance matrix is
\begin{equation}
\left(\begin{array}{cccc}
0.61\pm 0.02 & 0.00(4) \pm 0.02 & 0.29 \pm 0.02 & -0.01 \pm 0.02\\
0.00(4) \pm 0.02 & 0.61\pm 0.02 & 0.00(5)\pm 0.02 & -0.23 \pm 0.02\\
0.29 \pm 0.02 & 0.00(5)\pm 0.02 & 0.60\pm 0.02 & 0.00(2) \pm 0.02\\
-0.01\pm 0.02 & -0.23 \pm 0.02 & 0.00(2) \pm 0.02 & 0.60\pm 0.02
\end{array}\right),
\label{Eq:CM-exp}
\end{equation}
in which all elements consistent with zero are reported with the highest significant digit given in parentheses.
The matrix is written in the basis $(X_{a_{H,1}},Y_{a_{H,1}},X_{b_{V,-1}},Y_{b_{V,-1}})$
and the normalization is such that the shot-noise variance is $0.5$. This matrix
is consistent with a quantum state that has suffered 47\% of losses \cite{PRA2012}, in very good agreement with the
$0.52\pm0.03$ estimated collection efficiency of the detection set-up. This conclusion is based on the ``a priori''
assumption that the CM matrix must be symmetrical for the $a\leftrightarrow b$ exchange and that the small measured
asymmetries arise only from experimental fluctuations.
The same matrix is graphically reported in Fig.\ \ref{Fig:Matrix}.

In order to prove that this matrix witnesses genuine quantum correlations and represents an entangled CV bipartite Gaussian state, we used the Peres-Horodecki-Simon (PHS) \cite{Peres1996,Horodecki1997,Simon2000} and the Duan \cite{Duan2000} unseparability criteria. Using Eqs. (14)
\begin{equation}
n^{2}+m^{2}+2\left\vert c_{1}c_{2}\right\vert -4\left( nm-c_{1}^{2}\right)
\left( nm-c_{2}^{2}\right) \leq \frac{1}{4}
\label{Eq:PHS-criterion}
\end{equation}
and (20)
\begin{equation}
\sqrt{\left( 2n-1\right) \left( 2m-1\right) }-\left( c_{1}-c_{2}\right) \geq 0~.
\label{Eq:DUAN-criterion}
\end{equation}
of Ref.\ \cite{PRA2012} we found the measured state to be entangled with a high statistical significance ($\gtrsim 8 \sigma$) for both criteria. The above criteria for entanglement are necessary and sufficient only if the state is Gaussian as in this case. In our case, the left-hand-sides in Eqs. (\ref{Eq:PHS-criterion}) and  (\ref{Eq:DUAN-criterion}) are $0.51 \pm 0.03$ and $-0.31 \pm 0.04$, respectively.

\section{Physical state at the crystal}
\label{Sect:Physical_state}

As already mentioned, the state generated by the OPO is characterized by a single squeezing parameter, and this should constrain the relative CM to take the particular symmetrical form given in Eq.\ (\ref{Eq:twin-beam-pure-matrix}) with $n=m$ and $c_1=-c_2$.  Conversely, a non-symmetric process may generate a CM that depends on a pair of single-mode squeezing parameters, instead of one, allowing for a more general CM matrix with $n\neq m$ and $c_1 \neq -c_2$. Experimental imperfections and measurement fluctuations however always introduce small deviations from the above conditions, by introducing some degree of asymmetry. By reconstructing multiple CMs, we verified that these deviations are not systematic \cite{PRL2009,JOSAB2010,PRA2012,JOSAB2017}. We test the degree of symmetry of our as-measured CMs by checking that the following conditions are verified:
$n_{meas}-m_{meas}<2 (\Delta m_{meas}+ \Delta n_{meas})$ and $\left|c_{1,meas} \right|- \left|c_{2,meas} \right|<2 (\Delta c_{1,meas}+\Delta c_{2,meas})$, where in these equations $\Delta x$ denotes the standard deviation of the variable $x$. Only those matrices for which these conditions are satisfied are used for the subsequent analysis, while the others are discarded. Once a measured CM matrix is accepted, before using it for further analysis, we erase the unsymmetrical fluctuations by setting in Eq. (\ref{Eq:twin-beam-pure-matrix}) $c_1=-c_2=\left(\left|c_{1,meas} \right|+ \left|c_{2,meas} \right|\right)/2$ and $m=n=\left(m_{meas}+n_{meas}\right)/2$. The resulting matrix is called $\sigma_{symm}$.

At this stage we have two possible approaches for operatively retrieving the total loss that has affected the bipartite 
system so as to arrive at the state as it is generated inside the crystal. While it is clear that the state that would be, in the case, available for realising quantum information tasks that would exploit the OAM features is the one at the \textit{q-plate} output, the analysis of the ancestor state is an useful mean to certify that the generation, the manipulation and the detection methods are all consistent with their physical description.

We here discuss and compare the above mentioned approaches to obtain, by the measured CM, a measure of the total loss imparted by the system.

A first approach is to use the symmetric CM matrix itself to look for the loss level that allows one to obtain (by inverse transformation, see 
Ref.\ \cite{PRA2012} for details) a pure state at the crystal output and then compare it with the expected loss level.

A second approach, which in our opinion has the advantage of allowing for a double check of self-consistency, is offered by the measuring scheme itself.
The auxiliary modes $c$, $d$, $e$ and $f$ are individually single-mode squeezed states that suffered losses \cite{IJQI2007}. Tracing them back onto squeezed pure states yields a direct estimation of the loss level they suffered. Then, by averaging these values, we obtain a reliable estimate for the transmission factor $T$,
following the same approach employed in Ref.\ \cite{PRA2012}. Eventually, the CM of the ancestor bipartite pure two-mode squeezed entangled state can be calculated by the following expression \cite{Olivares2012}:
\begin{equation}
\sigma_0= \frac{1}{T} \left(\sigma_{symm} - \frac{1-T}{2}\; \mathbf{I}\right)
\end{equation}
where $\mathbf{I}$ is the 4x4 identity matrix.

Let us apply this second method to the experimental CM given in Eq.(\ref{Eq:CM-exp}).
We see that this is a matrix in the form of Eq. \ref{Eq:twin-beam-pure-matrix}, within the experimental errors. The measured values for $m=0.61\pm0.02$
and $n=0.60\pm0.02$, and $c_{1}=0.29\pm0.02$ and $c_{2}=-0.23\pm0.02$ are equal within two standard deviations, so they fulfil the above introduced symmetry condition.
We then erase the residual asymmetries, due to experimental imperfections, by setting the two pairs of elements equal to their average values (averages are calculated by using one more significant digit), obtaining
\begin{equation}
\sigma_{S}=\left(\begin{array}{cccc}
0.61 \pm 0.02 & 0 & 0.26 \pm 0.02 & 0\\
0 & 0.61 \pm 0.02 & 0 & -0.26 \pm 0.02\\
0.26 \pm 0.02 & 0 & 0.61 \pm 0.02 & 0\\
0 & -0.26 \pm 0.02 & 0 & 0.61 \pm 0.02
\end{array}\right)
\label{Eq:CM_exp_symm}
\end{equation}
The average losses calculated from single modes are $47\%$ ($T_{\sigma}=0.53$) so that the CM of the ancestor state matrix is
\begin{equation}
\sigma_{0}=\left(\begin{array}{cccc}
0.70 & 0 & 0.48 & 0\\
0 & 0.70 & 0 & -0.48\\
0.48 & 0 & 0.70 & 0\\
0 & -0.48 & 0 & 0.70
\end{array}\right)
\label{Eq:CM_pure_exp}
\end{equation}
The above matrix is physical and represents a twin-beam state whose purity is $0.99 \pm 0.03$.

\begin{figure}
	\begin{centering}
		\includegraphics[width=0.48\textwidth]{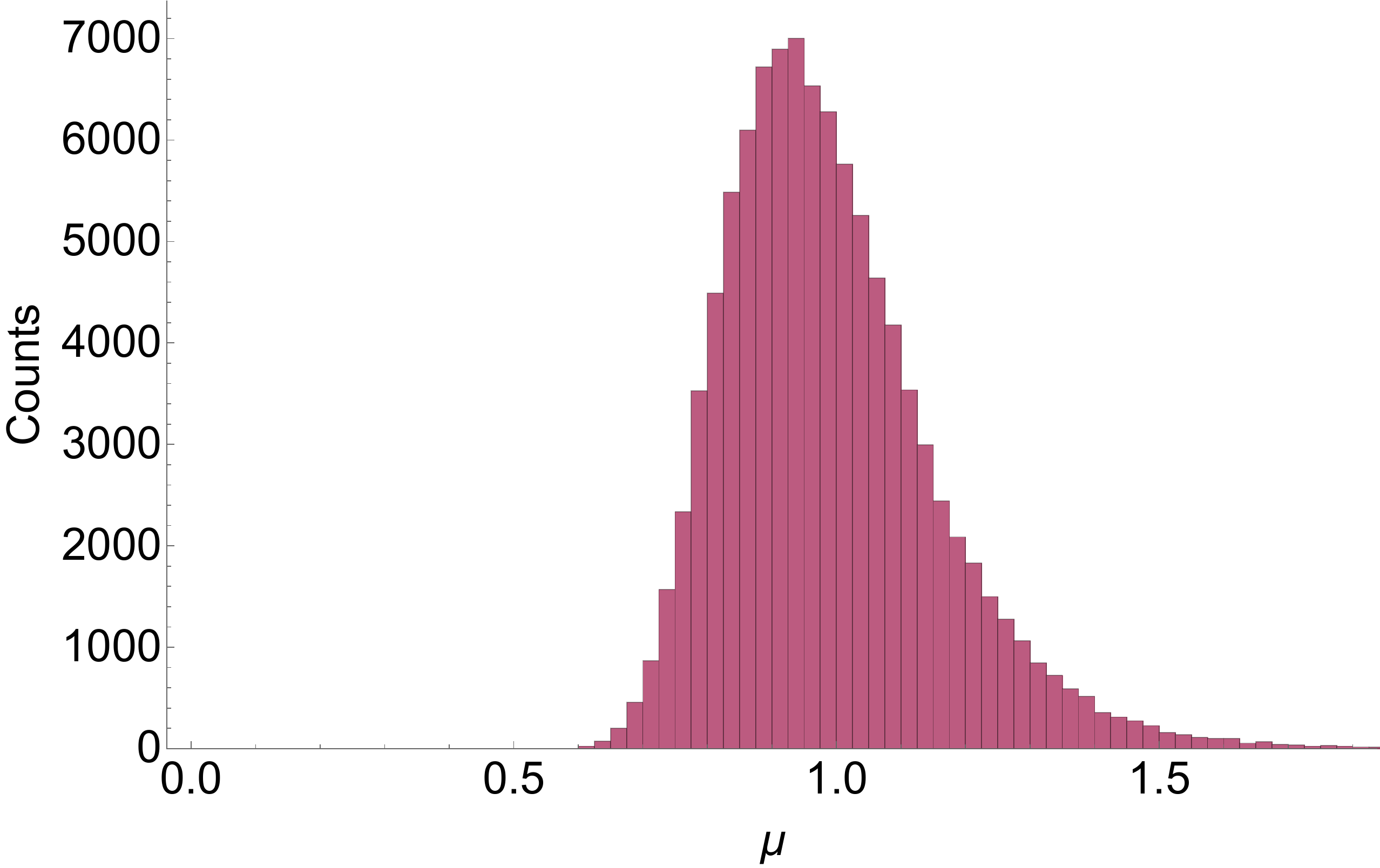}\caption{\label{Fig:Purity}Histogram of the purity $\mu$ (see Eq. \ref{Eq:purity}) for the reconstructed pure states obtained from the Monte Carlo statistical analysis. The average of the distribution is $0.99$ with a standard deviation of $0.03$. The distribution median is at $0.966$.}
	\end{centering}
\end{figure}

The error on the purity of the ancestor state is obtained by performing a MonteCarlo procedure in which each matrix element is randomly extracted $10^{5}$ times from a
Gaussian distribution centred on its experimental value and with standard deviation given by its experimental uncertainty. Each time the obtained covariance
matrix is symmetrized and the ancestor matrix calculated for $T=53\%$, according to the procedure described above. For each so calculated matrix we then extract the purity:
\begin{equation}
\mu=\frac{1}{4\sqrt{\mathrm{Det}(\sigma)}}
\label{Eq:purity}
\end{equation}
The set of $10^{5}$ values of the purities, whose histogram is given in Fig.\ \ref{Fig:Purity}, has been then statistically analyzed for calculating its mean and standard deviation.
We note that extracting from a Gaussian distribution the matrix elements can lead to unphysical results, i.e.\ $\mu>1$, but, being the distribution variance
sufficiently small with respect to the range of validity of $\mu$ itself, the use of a Gaussian distribution is still valid \cite{Olivares2012Metro}.
The average over $10^{5}$ samples give the mentioned result $\mu=0.99 \pm 0.03$.

These results confirm that the outlined procedure, although it may look complex and articulated, is consistent with both
the physics underlying the experiment and the particular experimental conditions. We also recall that the independently evaluated level of losses, taking into account detection efficiencies, cavity coupling efficiency, optical loss at the uncoated q-plate, and residual optical losses of the different optical components required for OAM 
manipulation and detection scheme, corresponds to $T=0.52\pm0.03$, which is fully compatible with $T_{\sigma}=0.53$ obtained from the above
data analysis.

\begin{figure}[tph]
	\centering\includegraphics[width=0.3\textwidth]{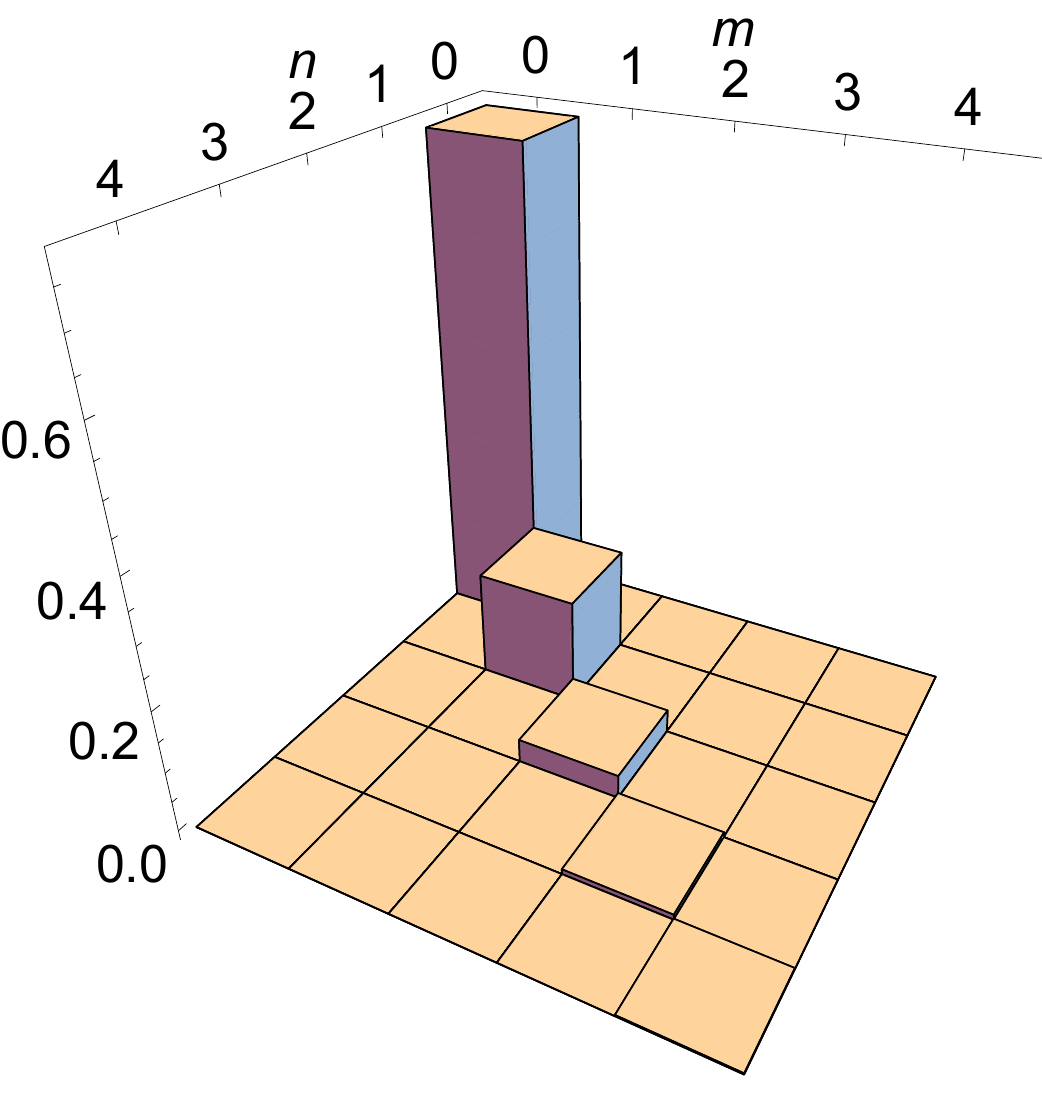}
	\caption{Joint photon number probability distribution $p(n;m)$ of the
		pure
		\textit{twin-beam} state
		generated inside the crystal. This state propagates all the way down to the homodyne detector and, after collection and detection losses,
		is represented by the CM given by Eq. (\ref{Eq:CM-exp}).}
	\label{Fig:Joint}
\end{figure}

From the measured CM, it is also possible to retrieve the joint photon number probability $p\left(n;m\right) $ (reported in Fig.\ \ref{Fig:Joint}) of the pure state generated in the crystal \cite{Olivares2012,JOSAB2010}.
It represents the probability of having exactly $m$ photons in mode $b$ and $n$ photons in mode $a$. For a pure two-modes squeezed state, only diagonal terms are nonzero. This is a clear signature that the system is in a \textit{twin-beam} state: every time a photon populates mode $a$ a twin one is certainly in mode $b$.

\section{Conclusions}
\label{Sect:Conclusions}

In conclusion, we have demonstrated the generation and complete experimental characterization of a bipartite continuous-variable entangled state endowed with non-zero OAM. In particular, we have completed the first experimental derivation of the full covariance matrix of the two helical modes, via measurement of the quadrature field correlations of various linear combinations of the generated modes. Although the experiment reported here concerns bipartite systems, the use of OAM has the potential for generating multipartite quantum states exploiting a hybrid discrete-continuous variable encoding, all within a single optical beam. This, in turn, could for example enable the multiplexing of multiple correlated quantum channels within a single optical channel. Indeed, a \textit{q--plate} acts as a beamsplitter in the infinite-dimensional OAM space thus opening the door to entangling many pairs of co-propagating orthogonal modes in a cascaded configuration.

\section*{Acknowledgement}
The authors thank S. Olivares and B. Piccirillo for granting the use of the routine for retrieving the joint photon number probability distribution and for the preparation of the q-plates used in the experiment, respectively. This work was partly supported by the European Union Horizon 2020 program, within the European Research Council (ERC) Grant No.\ 694683, PHOSPhOR.

\end{document}